\documentstyle[preprint,pre,aps]{revtex}

\newcounter{saveeqn}

\title{Application of the quantum spin glass theory \\
to image restoration}

\author{Jun-ichi Inoue} 
\address{Complex Systems Engineering, Graduate School of 
Engineering, \\
Hokkaido University, N13-W8, Kita-ku, Sapporo, 060-8628, Japan}
\date{\today}
\begin{document}
\maketitle
\thispagestyle{empty}
\begin{abstract}
Quantum fluctuation is introduced into 
the Markov random fields (MRF's) model for 
image restoration in 
the context of Bayesian approach. 
We investigate the dependence of the quantum 
fluctuation on the quality of 
BW image restoration by making use of statistical 
mechanics.    
We find that the maximum posterior marginal (MPM) estimate 
based on the quantum fluctuation gives a fine restoration  
in comparison with the maximum a posterior (MAP) estimate or 
the thermal fluctuation based MPM estimate.    
\end{abstract}
\mbox{}
\clearpage
\setcounter{page}{1} 
\section{Introduction}

Recently, the problems of information science were  
investigated from statistical mechanical point of view. 
Among them, the image restoration is one of the  
most suitable subjects. 
In the standard approach to the image restoration,  an estimate of the 
original image is given by maximizing 
a posterior probability distribution (the MAP estimate)
\cite{Geman}. 
In the context of statistical mechanics, 
this approach corresponds to finding 
the ground state configuration of the effective Hamiltonian 
for some spin system under the random fields. 
On the other hand, 
it is possible to construct another strategy 
to infer the original image using the 
thermal equilibrium state of the Hamiltonian. 
From the Bayesian statistical point of view, 
the {\it finite temperature restoration} 
coincides with maximizing 
a posterior marginal distribution 
(the MPM estimate \cite{Marroquin,PB}) and 
using this strategy, the error for each pixel 
may become smaller than that of the MAP estimate.   
As we use the average of each pixel (spin) over the 
Boltzmann-Gibbs distribution at a specific temperature, 
the thermal fluctuation should play an  
important role in the MPM estimate.  
Then, the temperature controls the shape of 
the distribution and if we 
choose the temperature appropriately, 
the sampling from the distribution 
generates the important 
configurations for a fine restoration.
Besides this hill-climbing mechanism by the 
thermal fluctuation, we may use another 
type of fluctuation, namely, 
the quantum fluctuation which leads to 
quantum tunneling between the states. 
If we use the sampling from the Boltzmann-Gibbs 
distribution based on the quantum fluctuation, it 
may be possible to 
obtain much more effective configurations for a good restoration. 
The idea of the MRF's model using the quantum fluctuation  
was recently proposed by Tanaka and Horiguchi \cite{TH}, however,  
they investigated the quantum fluctuation in the context of the  
optimization (the MAP estimate by the quantum fluctuation) 
and they used the ground state as the estimate of the 
original image. 
We would like to stress that we use 
the distribution based on the quantum fluctuation itself and 
the expectation value is used to infer the original image.     
It is highly non-trivial problem to 
investigate whether the MPM estimate based on the 
quantum fluctuation becomes better than the MAP estimate 
or the thermal fluctuation based MPM estimate.  

This is a basic concept of this paper.
This paper is organized as follows. 
In the next Sec. II,  
we explain our model system and 
the basic idea of our method in detail. 
In Sec. II, we also introduce the criterion of 
the restoration, that is, 
 the overlap between 
the original image and the result of the restoration.
In Sec. III,   
we introduce the infinite range 
model in order to obtain analytical results on 
the performance of the restoration, and 
calculate the overlap explicitly. 
In Sec. IV, 
we show that quantum Monte Carlo 
simulations in 2-dimension support our 
analytical results. 
In Sec. V, 
we introduce the iterative algorithm which is 
derived by mean-field approximation and 
apply this algorithm to image restoration 
for standard pictures. 
The last  Sec. VI is devoted 
to discussion about all results we obtain. 
In this section, we also mention the inequality 
which gives the upper bound of the overlap.
\section{Basic idea and formulation}
Let us suppose that 
the original image is represented by 
a configuration of Ising 
spins $\{ \xi \}\,
\equiv\,\{\xi_i|\xi_i=\pm 1 ; i=1,\cdots,N\}$ 
with probability $P_{s}
(\{ \xi \})$.
These images are sent 
through the noisy channel by 
the form of sequence $\{ \xi \}$. 
Then, we regard the output of the sequence 
$\{ \xi \}$ through the noisy channel as 
$\{ \tau \}$. 
The output probability for the case of the binary 
symmetric channel (BSC) is specified by the 
following form; 
\begin{eqnarray}
P_{\rm out}(
\{ \tau \}|\{ \xi \})=
\frac{1}{
(2{\cosh}{\beta}_{\tau})^{N}}
\,{\exp}\left(
{\beta}_{\tau}\sum_{i}{\tau}_{i}{\xi}_{i}
\right). 
\label{PoutBC}
\end{eqnarray}
We easily understand the relevance of this expression for  
the BSC; 
Lets suppose that each pixel $\xi_{i}$ changes its sign 
with probability $p_\tau$ and remains with $1-p_\tau$ 
during the transmission, 
that is,  
\begin{eqnarray}
P(\tau_i=-\xi_i|\xi_i) & = & p_\tau\,\equiv\,\frac{{\rm e}^{-\beta_\tau}}
{2\cosh \beta_\tau}  \\
P(\tau_i=\xi_i|\xi_i) & = & 1-p_\tau\,\equiv\,\frac{{\rm e}^{\beta_\tau}}
{2\cosh \beta_\tau}. 
\label{BSC_each_spin}
\end{eqnarray}
We easily see that there is a simple relation between 
flip probability $p_\tau$ and inverse temperature $\beta_\tau$ as 
${\exp}(2{\beta}_{\tau})=(1-p_{\tau})/p_{\tau}$. 
This is reason why we refer to this type of noise as 
{\it binary symmetric} channel.
Using the 
assumption that 
each pixel 
$\xi_i$ in 
the original image $\{\xi\}$ 
is corrupted 
independently 
(so-called {\it memory-less channel}), namely, 
$P(\{\tau\}|\{\xi\})=
\prod_i P(\tau_i|\xi_i)$, 
we obtain 
Eq. (\ref{PoutBC}).
This BSC is simply extended to the following 
Gaussian channel (GC)
\begin{eqnarray}
P_{\rm out}(\{ \tau \}|
\{ \xi \})  =  \frac{1}{(\sqrt{2\pi}\tau)^{N}}
\,{\exp}\left(
-\frac{1}{2{\tau}^{2}}
\sum_{i}
({\tau}_{i}-{\tau}_{0}{\xi}_{i})^{2}
\right).
\label{PoutGC}
\end{eqnarray}
where $\tau$ is a standard 
deviation of 
observable (corrupted pixel) $\tau_i$ 
from scaled original pixel  $\tau_0 \xi_i$.

Then, the posterior probability 
$P(\{ \sigma \}|
\{ \tau \})$, which is the probability 
that the source sequence is $\{ \sigma \}$ provided that 
the output is $\{ \tau \}$, 
leads to 
\begin{eqnarray}
P(\{ \sigma \}|\{ \tau \}) & = &  
\frac{P(\{ \tau \}
|\{ \sigma \})
P_{m}(\{ \sigma \})}
{\sum_{\sigma}P(\{ \tau \}
|\{ \sigma \})
P_{m}(\{ \sigma \})}
\label{Pcond}
\end{eqnarray}
by the Bayes theorem.
As we treat the BW image and the 
BSC (\ref{PoutBC}), 
a likelihood $P(\{\tau\}|\{\xi\})$ is appropriately written by 
\begin{eqnarray}
P(\{\tau\}|\{\xi\}) & \sim & 
{\exp}\left(h\sum_{i}\tau_i \sigma_i
\right).
\label{likelihood}
\end{eqnarray}
$P_{m}(\{ \sigma \})$ 
appearing in the 
Bayesian formula (\ref{Pcond}) 
is a model of the prior 
distribution $P_{s}(\{ \xi \})$ 
and we usually use the following type; 
\begin{eqnarray}
P_{m}(\{ \sigma \}) & \sim & {\exp}\left(
{{\beta}_{m}\sum_{<ij>}{\sigma}_{i}{\sigma}_{j}}
\right)
\label{Ferro2}
\end{eqnarray}
where $\sum_{<ij>}(\cdots)$ 
means the sum with respect to the nearest neighboring 
pixels and $\beta_{m}$ controls the smoothness of the picture according to 
our assumption. 
Substituting Eqs. (\ref{likelihood}) and 
(\ref{ferro}) into Eq. (\ref{Ppost}), we obtain  
the posterior probability $P(\{\sigma\}|\{\tau\})$ 
explicitly; 
\begin{eqnarray}
P(\{\sigma\}|\{\tau\}) & = & 
\frac{{\exp}\left(
\beta_m \sum_{<ij>}
\sigma_i \sigma_j +h\sum_i \tau_i \sigma_i
\right)}
{\sum_\sigma {\exp}\left(
\beta_m \sum_{<ij>}
\sigma_i \sigma_j +h\sum_i \tau_i \sigma_i
\right)}.
\label{Ppost}
\end{eqnarray}
In the framework of the MAP estimate,  
we regard a configuration $\{ \sigma \}$ which 
maximizes the posterior probability 
$P(\{ \sigma \}
|\{ \tau \})$ as an 
estimate of the 
original image $\{ \xi \}$. 
Obviously, this estimate $\{ \sigma \}$ 
corresponds to 
the ground state of 
the following effective Hamiltonian 
(the random field Ising model)  
\begin{eqnarray}
{\cal H}_{\rm eff} & = & -\beta_m 
\sum_{<ij>}\sigma_i \sigma_j -
h\sum_{i}{\tau}_{i} \sigma_{i}.
\label{effectH1}
\end{eqnarray}
Therefore, 
in the limit of $\beta_{m}/h\,{\rightarrow}\,\infty$,  
we expect that the original image should be complete BLACK picture or 
complete WHITE picture, whereas in the limit of 
$\beta_{m}/h\,{\rightarrow}\,0$, we 
assume that the original image should be 
identical to the observable  
$\{ \tau \}$ itself. 

On the other hand, 
in the framework of the MPM estimate, 
we maximize the following 
posterior marginal probability 
\begin{eqnarray}
P(\sigma_{i}|\{ \tau \})=
\frac{{\sum}_{\sigma\neq \sigma_{i}}
P(\{ \tau \}
|\{ \sigma \})
P_{m}(\{ \sigma \})}
{{\sum}_{\sigma}
P(\{ \tau \}
|\{ \sigma \})P_{m}(\{ \sigma \})}.
\label{marginal}
\end{eqnarray}
As we treat the case of 
BW image, the estimate of the $i$-th 
pixel should be 
given as 
\begin{eqnarray}
\mbox{} & \mbox{} &  {\rm sgn}
\left(
{\sum}_{\sigma_{i}=\pm 1}
{\sigma}_{i}P(\sigma_{i}|\{ \tau \})
\right)  \nonumber \\
\mbox{} & = & 
{\rm sgn}
\left(
\frac{{\sum}_{\sigma}\sigma_{i}
P(\{ \tau \}
|\{ \sigma \})
P_{m}(\{ \sigma \})}
{{\sum}_{\sigma}
P(\{ \tau \}
|\{ \sigma \})
P_{m}(\{ \sigma \})}
\right) \nonumber \\
\mbox{} & \equiv & {\rm sgn}(\langle {\sigma}_{i} \rangle_{h,\beta_{m}}) 
\end{eqnarray}
where $\langle \cdots \rangle_{h,\beta_{m}}$ 
means the average over 
the 
posterior probability 
Eq. (\ref{Ppost}).
Consequently, our problem is 
reduced to that of statistical mechanics which 
is described by the effective Hamiltonian 
Eq. (\ref{effectH1}). 
As the Hamiltonian Eq. (\ref{effectH1}) 
has lots of local minima 
due to the 
quenched disorder 
$\{ \tau \}$,  
in general, it is 
quite difficult to  
obtain the thermal equilibrium 
state which contributes to fine 
restoration without being 
trapped in a local minimum 
for a long time. 
In order to 
overcome this difficulty, 
we add the {\it quantum transverse field} \cite{CDS} 
\begin{eqnarray}
-{\Gamma}\sum_{i}\hat{\sigma}_{i}^{x}\,\equiv\,
\hat{\cal H}_{1}
\label{transverse}
\end{eqnarray}
to the effective Hamiltonian 
(\ref{effectH1}) as {\it quantum fluctuation}. 
In this expression, 
$\hat{\sigma}_{i}^{x}$ 
means the $x$-component 
of the Pauli matrix 
and $\Gamma$ controls 
the width of the quantum fluctuation. 
Intuitively, 
the term $\Gamma\, \hat{\sigma}_{i}^{x}$ 
is regarded 
as the {\it tunneling probability} 
between 
the eigenstates of the 
operator $\hat{\sigma}_{i}^{z}$ ($z$-component of the Pauli matrix), 
namely, 
$|{\sigma}_{i}^{z}=\pm 1>$. 
The tunneling probability between 
the states 
$|{\sigma}_{i}^{z}=\pm 1>$ leads to 
$|<{\sigma}_{i}^{z}=+1\,|\,\Gamma\, \hat{\sigma}_{i}^{x}
\,|\,{\sigma}_{i}^{z}=-1>|^{2}=\Gamma^{2}$.  
As the result, the term (\ref{transverse}) 
generates the superposition of 
the states $|\sigma_{i}^{z}=+1>$ (BLACK) 
and $|\sigma_{i}^{z}=-1>$ (WHITE). 
Using 
this {\it fuzzy} 
representation 
for each pixel, 
we may construct the 
algorithm which 
is robust for the choice of the 
hyper-parameters, especially, 
for the edge parts of a 
given picture. 

Our problem is now reduced to 
that of quantum statistical mechanics 
for the next effective Hamiltonian 
\begin{eqnarray}
\hat{\cal H}_{\rm eff} = 
-h\sum_{i}\tau_{i}\hat{\sigma}_{i}^{z}
-\beta_{m}\sum_{<ij>}
\hat{\sigma}_{i}^{z}
\hat{\sigma}_{j}^{z}-
\Gamma\,\sum_{i}\hat{\sigma}_{i}^{x}\, \equiv\,  
\hat{\cal H}_{0}+
\hat{\cal H}_{1} 
\label{effectH2}
\end{eqnarray}
where we defined $\hat{\cal H}_{1}\,\equiv\, 
\hat{\cal H}_{\rm eff}-\hat{\cal H}_{0}$.
Our main goal is to calculate the local 
magnetization $\langle\hat{\sigma}_{i}^{z} \rangle_{h,\beta_{m},\Gamma}$ 
of the system described by the above Hamiltonian, 
that is to say, 
\begin{eqnarray}
\langle \hat{\sigma}_i^{z} \rangle_{h,\beta_m,\Gamma} & \equiv & 
\frac{{\rm Tr}_{\sigma} \hat{\sigma}_{i}^{z}
{\exp}(-\hat{\cal H}_{\rm eff})}
{{\rm Tr}_{\sigma}
{\exp}(-\hat{\cal H}_{\rm eff})}
\label{aveQUANTUM}
\end{eqnarray}
and regard the quantity  
${\rm sgn}(\langle \hat{\sigma}_{i}^{z} \rangle_{h,\beta_{m},\Gamma})$ as 
an estimate of the original 
pixel ${\xi}_{i}$. 
Therefore, the averaged performance of 
our method is measured by the 
following {\it overlap} $M(h,\beta_{m},\Gamma)$ as 
\begin{eqnarray}
M(h,\beta_{m},\Gamma) & = & {\rm Tr}_{\{ \xi,
\tau \}}
P_{s}(\{ \xi \})
P_{\rm out}
(\{ \tau \}
|\{ \xi \})
{\xi}_{i}\,{\rm sgn}(\langle \hat{\sigma}_{i}^{z} 
\rangle_{h,\beta_{m},\Gamma}).
\label{overlap}
\end{eqnarray}     
Then, our main interests are 
summarized as follows.  
\begin{itemize}
\item 
Is it possible for us to  
use the quantum fluctuation in place of the thermal one ? 
\item
Does there exist a specific choice of 
$\Gamma$ which gives the optimal image 
restoration ?
\end{itemize}
Before we calculate the above 
overlap (\ref{overlap}),  
we may add the {\it parity check term},  which was recently 
introduced by Nishimori and Wong \cite{NW}, 
to the effective 
Hamiltonian (\ref{effectH1}). 
This parity check term is represented as  
$\beta_{J}\sum_{<ij>}J_{ij}\hat{\sigma}_{i}^{z}
\hat{\sigma}_{j}^{z}$, 
and we rewrite $\hat{\cal H}_{0}$ as 
\begin{eqnarray}
\hat{\cal H}_{0}=
-\beta_{J}\sum_{<ij>}J_{ij}\hat{\sigma}_{i}^{z}
\hat{\sigma}_{j}^{z}
-\beta_{m}\sum_{<ij>}\hat{\sigma}_{i}^{z}
\hat{\sigma}_{j}^{z}-
h\sum_{i}\tau_{i}\hat{\sigma}_{i}^{z}
\label{H02}
\end{eqnarray} 
 where 
$J_{ij}$ is the noisy version of the product of 
arbitrary two original pixels $\xi_{i}\xi_{j}$ and 
the output of this quantity through the noisy channel is 
given by  
\begin{eqnarray}
P_{\rm out}(\{ J \}|
\{ \xi \})=
\frac{1}{(2{\cosh}\beta_{r})^{N_{B}}}
\,{\exp}
\left(
\beta_{r}\sum_{<ij>}J_{ij}{\xi}_{i}{\xi}_{j}
\right)
\label{BSC2}
\end{eqnarray}
for the BSC and 
\begin{eqnarray}
P_{\rm out}(\{ J \}
|\{ \xi \}) =  
\frac{1}{(\sqrt{2\pi}J)^{N_{B}}}
\,{\exp}\,
\left(
-\frac{1}{2J^{2}}
\sum_{<ij>}(J_{ij}-J_{0}\xi_{i}\xi_{j})^{2}
\right)
\label{GC2}
\end{eqnarray}
for the GC, respectively. 
$N_{B}$ is the number of the terms appearing in the sum in 
Eq. (\ref{BSC2}) or Eq. (\ref{GC2}).  
Then, 
the effective Hamiltonian 
$\hat{\cal H}_{\rm eff}=\hat{\cal H}_{0}+\hat{\cal H}_{1}$ 
describes the thermo-dynamics of 
{\it quantum spin glass} \cite{CDS,BM} under 
random fields. 

In the next section, we introduce the rather artificial model, namely, 
the infinite range model 
in which spins in the system (\ref{effectH2}) are fully connected. 
\section{The Infinite Range Model }
In this section, 
we calculate the overlap (\ref{overlap}) 
explicitly using the infinite range version 
of the effective Hamiltonian (\ref{effectH2}).
We use the GC for the analysis of the 
infinite range model in this section 
and the the BSC for the quantum Monte Carlo simulations 
in the next Sec. IV, 
respectively.
However, 
these two channels can be treated by the following 
single form. 
\begin{eqnarray}
P_{\rm out}(\{ J \}|
\{ \tau \})=
\prod_{<ij>}
F_{r}(J_{ij})\prod_{<ij>}
F_{1}(\tau_{ij})\,{\exp}
\left(
\beta_{r}\sum_{<ij>}J_{ij}\xi_{i}\xi_{j}+
\beta_{\tau}\sum_{i}\tau_{i}\xi_{i}
\right)
\end{eqnarray}
with 
\begin{eqnarray}
F_{r}(J_{ij}) & = & 
\frac{1}{2{\cosh}\beta_{r}}
\left\{
{\delta}(J_{ij}-1)+{\delta}(J_{ij}+1)
\right\} \nonumber \\
\mbox{} F_{1}(\tau_{ij}) & = & 
\frac{1}{2{\cosh}\beta_{\tau}}
\left\{
{\delta}(\tau_{i}-1)+
{\delta}(\tau_{i}+1)
\right\}
\end{eqnarray}
for the BSC and with 
\begin{eqnarray}
F_{r}(J_{ij}) &  = & 
\frac{1}{\sqrt{2\pi J^{2}}}
{\exp}\,\left(
-\frac{1}{2J^{2}}
(J_{ij}^{2}+J_{0}^{2})
\right) \nonumber \\
\mbox{} F_{1}(\tau_{i}) & = & 
\frac{1}{\sqrt{2\pi \tau^{2}}}
{\exp}\,
\left(
-\frac{1}{2\tau^{2}}
(\tau_{i}^{2}+\tau_{0}^{2})
\right), 
\end{eqnarray}
for the GC, and we set $\beta_{J}=J_{0}/J^{2}, 
\beta_{\tau}=\tau_{0}/\tau^{2}$.

As the original image, 
we use the ferro-magnetic snapshot 
from the distribution 
\begin{eqnarray}
P_{s}(\{ \xi \})=
\frac{1}{{\cal Z}(\beta_{s})}
\,{\exp}
\left(
\frac{\beta_{s}}{N}\sum_{ij}
{\xi}_{i}{\xi}_{j}
\right), 
\label{ferro}
\end{eqnarray}
where $\sum_{ij}(\cdots)$ means   
the sum over all 
possible combinations of $(i,j)$ 
and we divided the argument of the exponential in Eq. (\ref{ferro}) 
by $N$ to take a proper thermo-dynamical limit 
 as Hamiltonian should be of order $N$. For the same reason, 
we should re-scale the terms appearing in 
 Eq. (\ref{effectH2}) as 
$\beta_{J}\sum_{<ij>}J_{ij}\hat{\sigma}_{i}^{z}
\hat{\sigma}_{j}^{z} \rightarrow 
(\beta_{J}/N)\sum_{ij}J_{ij}\hat{\sigma}_{i}^{z}
\hat{\sigma}_{j}^{z}$ and $\beta_{m}\sum_{<ij>}
\hat{\sigma}_{i}^{z}
\hat{\sigma}_{j}^{z} \rightarrow 
(\beta_{m}/N)\sum_{ij}\hat{\sigma}_{i}^{z}
\hat{\sigma}_{j}^{z}$ when we treat the infinite range model. 

It must be noted that 
$\hat{\cal H}_{0}$ and 
$\hat{\cal H}_{1}$ do not commute
with each other and we use the following the Trotter 
decomposition \cite{ST}
\begin{eqnarray}
{\cal Z}=\displaystyle{\lim_{P\rightarrow\infty}}
{\rm Tr}_{\sigma^{z}}\left(
{\rm e}^{-\frac{{\beta}{\cal H}_{0}}{P}}
{\rm e}^{-\frac{{\beta}{\cal H}_{1}}{P}}
\right)^{P}
\label{STD}
\end{eqnarray}
to calculate the partition function explicitly.  
In this formula, ${\cal H}_{0}$ and ${\cal H}_{1}$ are 
eigenvalues of the operators $\hat{\cal H}_{0}$ and 
$\hat{\cal H}_{1}$ with respect to 
the following eigenvector
\begin{eqnarray}
|\{\sigma_{k}^{z}\}> & = & \prod_{i=1}^{N}\bigotimes 
|{\sigma}_{i k}^{z}>\,\,\,\,\,\,(k=1,{\cdots},P)
\end{eqnarray} 
with  
\begin{eqnarray}
\hat{\sigma}_{i k}^{z}|\sigma_{i k}^{z}> & \equiv &  
{\sigma}_{i k}|{\sigma}_{i k}^{z}>.
\end{eqnarray}
$P$ means  
the Trotter number and we distinguish the different Trotter slices by 
the indices $k$.

Now we can calculate 
the partition function for the quantum 
spin system (\ref{H02})  
in terms of the 
corresponding classical spin system 
whose dimension increases by 1. 
Using the Trotter formula (the path integral formula) 
and well-known replica method \cite{SK}, namely, 
\begin{eqnarray}
[\ln {\cal Z}] & = & \displaystyle{\lim_{n{\rightarrow}\infty}}
\frac{[{\cal Z}^{n}]-1}{n},
\end{eqnarray}
we can obtain the overlap as a function of the 
macroscopic parameters $\beta_{m}$ and $\Gamma$ 
by making use of the saddle point method. 
The bracket $[\cdots]$ denotes 
the average over the distribution 
$P_{s}(\{ \xi \})
P_{\rm out}(\{ J \},
\{ \tau \}|
\{ \xi \})$.    

The standard replica calculations and 
saddle point method lead to 
the following coupled equations. 
\begin{eqnarray}
[{\xi}_{i}]=m_{0} & = & {\tanh}({\beta}_{0}m_{0}) 
\label{m01} \\
\mbox{}[\langle {\sigma}_{iK}^{\alpha} \rangle_{h,\beta_{m},\Gamma}]=m & = & 
\frac{{\rm Tr}_{\xi}\,{\rm e}^{{\beta}_{s}m_{0}{\xi}}}
{2\,{\cosh}({\beta}_{s}m_{0})}
\int_{-\infty}^{\infty}\!\!\!
Du\, {\Omega}^{-1}
\int_{-\infty}^{\infty}\!\!\!
D{\omega}\, {\Phi}y^{-1}\, {\sinh}y 
\label{m1} \\
\mbox{}[{\xi}_{i}\langle {\sigma}_{iK}^{\alpha} 
\rangle_{h,\beta_{m},\Gamma}]= t & = & 
\frac{{\rm Tr}_{\xi}\,{\rm e}^{{\beta}_{s}m_{0}{\xi}}}
{2\,{\cosh}({\beta}_{s}m_{0})}
\int_{-\infty}^{\infty}\!\!\!
Du\, {\Omega}^{-1}
\int_{-\infty}^{\infty}\!\!\!
D{\omega}\, {\xi}\, {\Phi}y^{-1}\, {\sinh}y 
\label{t1} \\
\mbox{}[\langle ({\sigma}_{iK}^{\alpha})^{2} 
\rangle_{h,\beta_{m},\Gamma}
]=Q & = & 
\frac{{\rm Tr}_{\xi}\,{\rm e}^{{\beta}_{s}m_{0}{\xi}}}
{2\,{\cosh}({\beta}_{s}m_{0})}
\int_{-\infty}^{\infty}\!\!\!
Du
\left[
{\Omega}^{-1}
\int_{-\infty}^{\infty}\!\!\!
D{\omega}\, {\Phi}y^{-1}\, {\sinh}y
\right]^{2} 
\label{Q1} \\
\mbox{}[\langle {\sigma}_{iK}^{\alpha}{\sigma}_{iL}^{\alpha} 
\rangle_{h,\beta_{m},\Gamma}]=S & = & 
\frac{{\rm Tr}_{\xi}\,
{\rm e}^{{\beta}_{s}m_{0}{\xi}}}
{2\,{\cosh}({\beta}_{s}m_{0})}
\int_{-\infty}^{\infty}\!\!\!
Du\, {\Omega}^{-1}
{\Biggr [}
\int_{-\infty}^{\infty}\!\!\!
D{\omega}\, {\Phi}^{2}y^{-2}
{\cosh}y \nonumber \\
\mbox{} & + & {\Gamma}^{2}
\int_{-\infty}^{\infty}\!\!\!
D{\omega}\, y^{-3}{\sinh}y
{\Biggr ]},
\label{S1}
\end{eqnarray}
where $\langle \cdots \rangle_{h,\beta_m,\Gamma}$ 
means the average by the 
posterior probability 
using the same way as Eq. (\ref{aveQUANTUM}).  
$Du$ or $Dy$ means Gaussian integral measure 
$Du\,\equiv\,du\,{\rm e}^{-u^2/2}/\sqrt{2\pi}$. 
In order to obtain the above saddle point equations, we 
used the replica symmetric and the static approximation, that is, 
\begin{eqnarray}
t_{K} & = & t 
\label{RSandSTtIR}\\
S_{\alpha}(KL) & = & S (K\,\,{\neq}\,\,L),\,\,\,1(K=L)
\label{RSandSTSIR} \\
Q_{\alpha\beta} & = & Q. 
\label{RSandSTQIR}
\end{eqnarray}
We also defined 
functions ${\Phi}$, $y$ and ${\Omega}$ as  
\begin{eqnarray}
{\Phi} & \,{\equiv}\, &  
u\sqrt{(\tau h)^{2}+Q(J\beta_{J})^{2}}
+J{\beta}{\omega}\sqrt{S-Q}
+({\tau}_{0}h+J_{0}{\beta}_{J} t){\xi}
+{\beta}_{m}m 
\label{Phi1} \\
y & \,{\equiv}\, & 
\sqrt{{\Phi}^{2}+\Gamma^{2}} 
\label{y1} \\
{\Omega} & \,{\equiv}\, & 
\int_{-\infty}^{\infty}\!\!\!
D{\omega}\,\,{\cosh}y.
\label{Omega1}
\end{eqnarray} 
Then the overlap which is a measure of retrieval quality is 
calculated explicitly as 
\begin{eqnarray}
\mbox{} & \mbox{} & [{\xi}_{i}\, {\rm sgn}(\langle {\sigma}_{iK}^{\alpha} 
\rangle_{h,\beta_{d},\Gamma})]  =  M =
\frac{{\rm Tr}_{\xi}\,{\xi}{\rm e}^{{\beta}_{s}m_{0}{\xi}}}
{2\,{\cosh}({\beta}_{s}m_{0})}
\int_{-\infty}^{\infty}\!\!\!
Du 
\int_{-\infty}^{\infty}\!\!\!
Dw
\nonumber \\
\mbox{} & {\times} & {\rm sgn}
\left[
u\sqrt{(\tau h)^{2}+Q(J\beta_{J})^{2}}
+({\tau}_{0}h+J_{0}{\beta}_{J}t){\xi}+{\beta}_{m}m +J{\beta}_{J}w\sqrt{S-Q}
\right]
\label{overlap2}
\end{eqnarray}
where the above overlap $M$ depends on 
$\Gamma$ through $m$ (\ref{m1}). 

We first consider the case of 
$\beta_{J}=0$, that is 
to say, the conventional 
image restoration. 
We choose a snapshot 
from the distribution (\ref{ferro}) 
at source temperature $T_{s}=0.9$. 
According to 
Nishimori and Wong \cite{NW}, 
we fix the ratio 
$h/\beta_{m}$ and adjust $\beta_{m}(=1/T_{m})$ 
as a parameter for 
simulated annealing \cite{Kirk} and controls $\Gamma$ as a 
quantum fluctuation. 
If we set $\Gamma=0$, the lines of $M(T_{m}, \Gamma=0)$ should 
be identical to the results by 
the {\it thermal}  MPM estimate \cite{NW}. 
On the other hand, 
if we choose $T_{m}=0$ and $\Gamma=0$, 
the resultant line $M(T_m=0, \Gamma)$ 
represents the performance of the 
{\it quantum} MAP estimate. 
We should draw attention to 
the fact that 
the quantum fluctuation vanishes 
at $\Gamma=0$. 
In practical applications of 
the {\it quantum annealing} \cite{KN}
 based on quantum Monte Carlo 
simulations, 
we should 
reduce $\Gamma$ from $\Gamma > 0$ to 
$\Gamma=0$ during Monte Carlo updates. 
However, the resultant 
performance obtained here 
is calculated analytically provided that the 
system reaches its equilibrium state. 
Therefore, 
we can regard the result 
$M(T_{m}=0,\Gamma=0)$ as a performance 
when $\Gamma$ is decreased slowly enough. 

In FIG. \ref{fig1}, 
we set the ratio $h/\beta_{m}$ to its 
optimal value $\beta_{\tau}/\beta_{s}=0.9$ and 
plot the overlap $M(T_{m},\Gamma)$ for 
the case of $\Gamma=0,0.5$ and $\Gamma=1.0$. 
Obviously, for the case of 
$\Gamma=0$, 
the maximum 
is obtained at a specific temperature 
$T_{m}=0.9 (=T_{s})$ \cite{NW}. 
However, 
if we add a finite quantum fluctuation, 
the optimal temperature $T_{m}$ is 
shifted to the low temperature region. 

In FIG. \ref{fig2}, 
we plot $M(T_m,\Gamma)$ for the case of 
$T_{m}=0,0.1,0.9$ 
with the fixed optimal ratio 
$h/\beta_{m}=0.9$. 
This figure shows that 
if we set the parameters $h, \beta_m$ 
to their optimal value in  
the thermal MPM estimate, 
the quantum fluctuation 
added to the system destroys 
the recovered image (see the lines $M(T_m,\Gamma)$ 
for the case of $T_{m}=0.9$).
Therefore, we may say that 
it is impossible to 
choose all parameters $h,\beta_{m}$ and $\Gamma$ 
so as to obtain the overlap which is larger 
than $M_{\rm max}\,\equiv\, M(T_m=0.9,\Gamma=0)$. 
This fact is also shown by 
$3$-dimensional plot $M(T_m,\Gamma)$ in 
FIG. \ref{fig3}. 

Although, we found that a finite 
$\Gamma$ does not give the 
absolute maximum of the overlap, 
the {\it quantum} MPM estimate 
$M(T_m=0,\Gamma >0)$ has 
another kind of advantages. 
As FIG. \ref{fig3} indicates, 
the overlap of the the quantum  
MPM estimate is almost flat 
in comparison with 
$M(T_m=0.1,\Gamma>0)$ or $M(T_m=0.9,\Gamma>0)$. 
This is a desirable property from 
practical point of view. 
This is because the estimation of the 
hyper-parameters 
is one of the crucial 
problems to infer 
the original image, and in general, 
it is difficult to estimate them beforehand. 
Therefore, 
this robustness 
for hyper-parameter selection is 
a desirable property. 
We also see this property in FIG. \ref{fig3}. 

As we already mentioned, the 
 overlap at $T_{m}=0$ and $\Gamma=0$ 
corresponds to the result which 
is obtained by quantum annealing \cite{KN}, 
that is to say, 
the quantum MAP estimate. 
We see that 
the result of the quantum MPM estimate 
is slightly 
better than that of the quantum 
MAP estimate. 

We next show the effect of the parity check term. 
In FIG. \ref{fig4}, 
we set $T_{m}=T_{s}=0.9, h=1.0$ and 
$J_{0}=J=1.0$ and plot 
the overlap as a function of 
$\beta_{J}$ for several values of $\Gamma$. 
We see that the performance 
of the restoration is improved 
by introducing the parity check 
term which has much information 
about the local structure of the 
original image. 

In the next section, we check the usefulness of this method 
 in terms of quantum Monte Carlo simulation.
\section{Quantum Monte Carlo Simulation}
In this section, 
Monte Carlo simulations in realistic $2$-dimension are carried out 
in order to check the practical usefulness of our method.   
We use the {\it standard pictures} which are provided on the 
web site \cite{Standard} as the 
original image, instead of the Ising snapshots. 
In order to sampling the important points which 
contribute to the local magnetization 
$\langle \hat{\sigma}_{i}^{z} \rangle$,  
we use the quantum Monte Carlo method which was 
proposed by Suzuki \cite{ST}.
As we mentioned in the previous sections, 
we can treat 
the $d$-dimensional quantum system 
as $(d+1)$-dimensional classical system by 
the Trotter decomposition \cite{ST}.
In this sense, the transition probability of the 
Metropolis algorithm leads to 
\begin{eqnarray}
P(\mbox{\boldmath $\sigma$}\,{\rightarrow}\,\mbox{\boldmath $\sigma$}^{'})=
{\min}
\left[
1,{\exp}(-(E({\mbox{\boldmath $\sigma$}}^{'})-
E({\mbox{\boldmath $\sigma$}})))
\right]
\label{trans}
\end{eqnarray}
where $E(\mbox{\boldmath $\sigma$})$ 
is energy of the classical spin system 
in $(d+1)$-dimension (in the present case, 
$(2+1)=3$-dimension) as follows. 
\begin{eqnarray}
E(\mbox{\boldmath $\sigma$}) & {\equiv} &    
-\frac{\beta_{m}}{P}\sum_{ijk}
[
{\sigma}_{i,j,k}{\sigma}_{i+1,j,k}+
{\sigma}_{i,j,k}{\sigma}_{i-1,j,k} 
+   
{\sigma}_{i,j,k}{\sigma}_{i,j+1,k}+
{\sigma}_{i,j,k}{\sigma}_{i,j-1,k}
] \nonumber \\
\mbox{} & - &  \frac{h}{P}\sum_{ijk}{\tau}_{i,j}
{\sigma}_{i,j,k} 
-  B\sum_{ijk}{\sigma}_{i,j,k}{\sigma}_{i,j,k+1}
\label{energy}
\end{eqnarray}
where we defined  $B\,\equiv\,{\ln}\,{\cosh}(\Gamma/P)$.
The transition probability  Eq. (\ref{trans}) with 
Eq. (\ref{energy}) generates 
the Boltzmann-Gibbs distribution 
asymptotically and 
using the importance sampling from the distribution, 
we can calculate the expectation value of the $i$-th  
spin $\hat{\sigma}_{i}^{z}$, 
namely, $\langle \hat{\sigma}_{i}^{z} \rangle_{h,\beta_{m},\Gamma}$, 
and using this result, 
we obtain an estimate of the $i$-th pixel 
of the original image 
as ${\rm sgn}(\langle \hat{\sigma}_{i}^{z} 
\rangle_{h,\beta_{m},\Gamma})$. 
We show the results 
in FIG.s \ref{fig5} and \ref{fig6}. 
From these Figures, we see that 
there exists the optimal value of the 
transverse field $\Gamma$ . 
In FIG.s \ref{fig7} and \ref{fig8}, 
we display the results by quantum Monte Carlo simulations 
when we add the parity check term for the 
parameter sets $\Gamma=2.0, h=1.0$ and $\beta_{m}=0.5$. 
We see that the resultant pictures using the parity check term are 
almost perfect (see $\beta_{J}=1.0$ and $1.5$).
\section{Mean-field algorithm}
In the previous sections, we 
see that the quantum fluctuation works effectively 
on image restoration problems in the 
sense that the quantum fluctuation 
suppress the error of the hyper-parameter's 
estimation in the Markov random fields model.
In addition, by making use of the 
quantum Monte Carlo simulations, 
we could apply it to  
the image restoration of the 2-dimensional 
standard pictures. 
However, 
in order to carry out the simulations, 
it takes quite long time to 
obtain the average 
$\langle \hat{\sigma}_{i}^{z} \rangle_{h,\beta_{m},\Gamma}$ 
and it is not suitable for practical situations.

In this section, in order to overcome 
this computational time intractability, 
we derive the 
iterative algorithm based on the 
mean-field approximation. 
This algorithm  
shows fast convergence to the approximate solution. 

Within the mean-field 
approximation, we rewrite the density 
matrix 
$\hat{\rho}={\rm e}^{-\hat{\cal H}_{\rm eff}}/{\cal Z}$ 
for 2-dimensional version of the 
effective Hamiltonian 
$\hat{\cal H}_{\rm eff}$ as 
\begin{eqnarray}
\hat{\rho}\,\simeq\, \prod_{ij}{\bigotimes}\hat{\rho}_{ij}, 
\end{eqnarray}
where we defined $\hat{\rho}_{ij}$ as 
\begin{eqnarray}
\hat{\rho}_{ij}=
\sum_{n=1}^{2}|\sigma_{ij}(n)>{\rm e}^{-\lambda_{ij}(n)}
<\sigma_{ij}(n)|
\end{eqnarray}
with 
\begin{eqnarray}
{\cal Z}_{ij}={\rm e}^{-\lambda_{ij}(1)}
+{\rm e}^{-\lambda_{ij}(2)}. 
\end{eqnarray}
In the above expressions, 
$\lambda_{ij}(n),\,\,n=1,2$ 
means  eigenvalues of the $2{\times}2$ matrix 
$\hat{\mbox{\boldmath $H$}}_{ij}$ (
$[\hat{\mbox{\boldmath $H$}}_{ij}]_{11}=H_{ij}^{(+)}, 
[\hat{\mbox{\boldmath $H$}}_{ij}]_{22}=H_{ij}^{(-)}, 
[\hat{\mbox{\boldmath $H$}}_{ij}]_{12}=[
\hat{\mbox{\boldmath $H$}}_{ij}]_{21}=-\Gamma$) 
and $H_{ij}^{(\pm)}$ is defined by  
\begin{eqnarray}
H_{ij}^{(+)} & = & 
-(\tau_{ij}+Jm_{i+1,j}^{(t)}
+Jm_{i-1,j}^{(t)}+
Jm_{i,j+1}^{(t)}+Jm_{i,j-1}^{(t)}) \nonumber \\
\mbox{} & = & -H_{ij}^{(-)} \equiv \alpha, \\
J & \equiv & \frac{\beta_m}{h}.
\end{eqnarray}
Using this 
decoupled density matrix, the local 
magnetization at a site $(i,j)$, 
namely, $m_{ij}^{(t+1)}$ 
leads to 
\begin{eqnarray}
m_{ij}^{(t+1)} & = & {\rm Tr}[\sigma_{ij}^{z}
\hat{\rho}_{ij}] \nonumber \\
\mbox{} & = & 
\frac{{\rm e}^{\sqrt{\alpha^{2}+\Gamma^{2}}}}
{2{\cosh}(\sqrt{\alpha^{2}+\Gamma^{2}})}
\left[
\frac{(\alpha+\sqrt{\alpha^{2}+\Gamma^{2}})^{2}-
\Gamma^{2}}
{(\alpha + \sqrt{\alpha^{2}+\Gamma^{2}})^{2}+\Gamma^{2}}
\right] \nonumber \\
\mbox{} & + & 
\frac{{\rm e}^{-\sqrt{\alpha^{2}+\Gamma^{2}}}}
{2{\cosh}(\sqrt{\alpha^{2}+\Gamma^{2}})}
\left[
\frac{(\alpha-\sqrt{\alpha^{2}+\Gamma^{2}})^{2}-
\Gamma^{2}}
{(\alpha-\sqrt{\alpha^{2}+\Gamma^{2}})^{2}+\Gamma^{2}}
\right].
\label{iteration}
\end{eqnarray}
For this local magnetization (\ref{iteration}), 
the estimate of the pixel $\xi_{ij}$ 
is obtained as ${\rm sgn}[m_{ij}]$.

We solve the mean-field equations (\ref{iteration}) with 
respect to $m_{ij}$ until 
the condition 
\begin{eqnarray}
{\varepsilon}_{ij}\,\equiv\,
|m_{ij}^{(t+1)}-m_{ij}^{(t)}|<10^{-5}
\end{eqnarray}
holds for all  pixels $\{i,j\}$. 
We show its performance in 
 FIG. \ref{fig9} and TABLE \ref{table1}. 
From TABLE \ref{table1}, we see that 
if we introduce appropriate quantum fluctuation, 
the performance is remarkably improved, and in addition, 
the speed of the convergence 
becomes much faster. 
However, if we add the quantum fluctuation 
too much, the fluctuation destroys the 
recovered image. 
We also see that the optimal value of $\Gamma$ 
exists around $\Gamma \sim 1.6$.
\section{Summary and Discussion} 
In this paper, we investigated to what extent 
the quantum fluctuation works effectively 
on image restoration. 
For this purpose, we introduced an analytically 
solvable model, 
that is,  the infinite range version of the 
MRF's model. 
We applied the technique of statistical 
mechanics to this model and derived the overlap 
explicitly. 
We found that the quantum fluctuation 
improves the quality of the 
image restoration dramatically at a low temperature region. 
In this sense, the error of the estimation for the 
hyper-parameters $\beta_{m}, h$  
can be suppressed by the quantum fluctuation. 

However, we also found that 
the maximum value of the overlap $M_{\rm max}^{\rm (qunatum)}$ 
never exceeds that of 
the classical Ising case $M_{\rm max}^{\rm (thermal)}$.
We may show this fact by the following arguments;
First of all,
 the upper bound of the overlap for the 
classical system is given by setting  
$h=\beta_{\tau}, P_{s}=P_{m}$, that is, 
\begin{eqnarray}
M_{\rm max}^{{\rm (thermal)}}
(\beta_{\tau},P_{s}) & = & 
{\rm Tr}_{\{\tau,\xi\}}
\xi_{i}{\rm e}^{\beta_{\tau}\sum_{i}\tau_{i}\xi_{i}}P(\{ \xi\} )\,
{\rm sgn}[{\rm Tr}_{\sigma}
\sigma_{i}{\rm e}^{\beta_{\tau}\sum_{i}\tau_{i}\sigma_{i}}
P_{m}(\{ \sigma \})] \nonumber \\
\mbox{} & =  & 
{\rm Tr}_{\{\tau,\xi\}}
{\xi}_{i}{\rm e}^{\beta_{\tau}\sum_{i}\tau_{i}\xi_{i}}
P(\{ \xi\} ){\cdot}
\frac{{\rm Tr}_{\sigma}
\sigma_{i}{\rm e}^{\beta_{\tau}\sum_{i}\tau_{i}
\sigma_{i}}P_{m}(\{ \sigma\} )}
{|{\rm Tr}_{\sigma}
\sigma_{i}{\rm e}^{\beta_{\tau}\sum_{i}\tau_{i}\sigma_{i}}
P_{m}(\{ \sigma\} )|} \nonumber \\
\mbox{} & = & {\rm Tr}_{\tau}
|{\rm Tr}_{\sigma}
\sigma_{i}{\rm e}^{\beta_{\tau}\sum_{i}\tau_{i}\sigma_{i}}
P_{m}(\{ \sigma\} )|. 
\end{eqnarray}
For the quantum system, the 
overlap is bounded by this maximum value $M_{\rm max}^{{\rm (classical)}}$ 
as 
\begin{eqnarray}
M^{{\rm (quantum)}}(h,P_{m},\Gamma) & = & 
{\rm Tr}_{\{\tau,\xi\}}
{\xi}_{i}{\rm e}^{\beta_{\tau}\sum_{i}\tau_{i}\xi_{i}}
P(\{ \xi \})\,{\rm sgn}
[{\rm Tr}_{\hat{\sigma}}
\hat{\sigma}_{i}^{z}\,
{\rm e}^{h\sum_{i}{\tau}_{i}\hat{\sigma}_{i}^{z} 
+{\Gamma}\sum_{i}\hat{\sigma}_{i}^{x}}
P_{m}(\hat{\sigma}^{z})] \nonumber \\
\mbox{} & \,\leq\,  & |
{\rm Tr}_{\{\tau,\xi\}}
{\xi}_{i}
{\rm e}^{\beta_{\tau}\sum_{i}\tau_{i}\xi_{i}}
P(\{ \xi\} )
{\rm sgn}[
{\rm Tr}_{\hat{\sigma}}
\hat{\sigma}_{i}^{z}
{\rm e}^{h\sum_{i}\tau_{i}\hat{\sigma}_{i}^{z}+
\Gamma \sum_{i}\hat{\sigma}_{i}^{x}}
P_{m}(\{ \hat{\sigma}_{i}^{z}\} )]| \nonumber \\
\mbox{} & =  & 
{\rm Tr}_{\tau}|{\rm Tr}_{\xi}
{\xi}_{i}{\rm e}^{\beta_{\tau}\sum_{i}\tau_{i}\xi_{i}}
P(\{ \xi\} )|=M_{\rm max}^{{\rm (thermal)}}.
\end{eqnarray}
We can see this inequality 
more directly as follows.   
\begin{eqnarray}
{\rm Tr}_{\tau}|
{\rm Tr}_{\xi}\xi_{i}
{\rm e}^{\beta_{\tau}\sum_{i}\tau_{i}\xi_{i}}
P(\{ \xi\} )| &\, {\geq}\, & 
{\rm Tr}_{\{\tau,\xi\}}
{\xi}_{i}{\rm e}^{\beta_{\tau}\sum_{i}\tau_{i}\xi_{i}}
P(\{ \xi\} ){\cdot}
\frac{{\rm Tr}_{\hat{\sigma}}
\hat{\sigma}_{i}^{z}\,
{\rm e}^{h\sum_{i}{\tau}_{i}\hat{\sigma}_{i}^{z} 
+{\Gamma}\sum_{i}\hat{\sigma}_{i}^{x}}
P_{m}(\{ \hat{\sigma}^{z}\} )}
{|{\rm Tr}_{\hat{\sigma}}
\hat{\sigma}_{i}^{z}\,
{\rm e}^{h\sum_{i}{\tau}_{i}\hat{\sigma}_{i}^{z} 
+{\Gamma}\sum_{i}\hat{\sigma}_{i}^{x}}
P_{m}(\{ \hat{\sigma}^{z}\} )|} \nonumber \\
\mbox{} & = & 
{\rm Tr}_{\{\tau,\xi\}}
{\xi}_{i}{\rm e}^{\beta_{\tau}\sum_{i}\tau_{i}\xi_{i}}
P(\{ \xi\} )
\,{\rm sgn}[{\rm Tr}_{\hat{\sigma}}
\hat{\sigma}_{i}^{z}\,
{\rm e}^{h\sum_{i}{\tau}_{i}\hat{\sigma}_{i}^{z} 
+{\Gamma}\sum_{i}\hat{\sigma}_{i}^{x}}
P_{m}(\{ \hat{\sigma}^{z}\} )] \nonumber \\
\mbox{} & = & M^{{\rm (quantum)}}(h,P_{m},\Gamma),
\label{inequality3} 
\end{eqnarray}
where  the identity ${\rm sgn}(x)=x/|x|$  
was used. 
We should notice that in the left hand side of the 
above inequality (\ref{inequality3}), the arguments of the trace 
w. r. t.  $\tau$ always take positive values, while 
in the right hand side, they can be negative.  

In order to check the usefulness of the method, 
we carried out 
quantum Monte Carlo 
simulations in realistic $2$-dimension. 
We found that the results by the simulation 
support qualitative behavior of the analytical 
expressions for overlap. 

We introduced the iterative algorithm in terms of 
the mean-field approximation 
and applied it to image restoration 
of the standard pictures. 
We found that the quantum fluctuation 
suppress the error of the 
hyper-parameter estimation. 
In addition, we found that 
the speed of the convergence to 
the solution is accelerated by the 
quantum fluctuation.

From all results obtained in this paper, 
we concluded that 
the quantum fluctuation 
turns out to enhance tolerance against 
uncertainties in hyper-parameter 
estimation. 
However, if much higher quantities of 
restoration are required, we must 
estimate those parameters using some 
methods. 
One of the strategies for this 
purpose is selecting the parameters 
$\beta_{m},h$ and $\Gamma$ 
which maximize a {\it marginal 
likelihood}. 
By making use of the infinite 
range model, the 
usefulness of this method can be 
evaluated. 
The details of the analysis 
will be reported in forth coming paper.

Of course, the application of this strategy 
to the restoration of  
gray-scaled image \cite{MI,IM} 
will be considered as an important future problem. 

The author acknowledges H. Nishimori  
for fruitful 
discussions and useful comments. 
He also thanks K. Tanaka 
for kind tutorial on the 
theory of image restoration 
and drawing his attention to 
reference \cite{TH}. 
He acknowledges  D. Bolle' , A. C. C. Coolen,  D. M. Carlucci, T. Horiguchi,
P. Sollich and  K. Y. M. Wong  for valuable discussions. 
The author thanks Department of Physics, Tokyo Institute of Technology 
and Department of Mathematics,  
Kings College, University of London for hospitality. 

This work was partially   
supported by the Ministry of Education, 
Science, Sports and Culture, 
Grant-in-Aid for 
Encouragement of Young Scientists, 
No. 11740225, 1999-2000 and 
also supported by the 
collaboration program 
between Royal Society and 
Japanese Physical Society.  

\begin{figure}
\caption{The overlap $M$ as 
a function of $T_{m}$ for several 
values of  $\Gamma$.  
We set the system parameters as 
$T_{s}=0.9, \tau_{0}=\tau=1.0$ and 
$h/\beta_{m}=0.9=\beta_\tau/\beta_{s}$. 
For $\Gamma=0$ case, 
the optimal temperature $T_{m}$ coincides with 
the source temperature $T_{s}=0.9$. 
As the quantum fluctuation $\Gamma$ increases, 
the optimal temperature is 
shifted to a low temperature 
region.  However, the maximum value of 
the overlap does not change.}
\label{fig1} 
\end{figure}
\begin{figure}
\caption{The overlap $M$ as a function of $\Gamma$ 
for several values of $T_{m}$.  
We set the system parameters as 
$T_{s}=0.9, \tau_{0}=\tau=1.0$ and 
$h/\beta_{m}=0.9=\beta_\tau/\beta_{s}$. 
The overlap at $T_{m}=0$ and $\Gamma=0$ corresponds to 
the result by quantum annealing. 
The quantum MPM estimate works 
effectively at a low temperature 
region and the results are robust for the choice of $\Gamma$.
}
\label{fig2}
\end{figure}
\begin{figure}
\caption{The overlap $M$ as a function of the 
quantum fluctuation $\Gamma$ and 
temperature $T_{m}$.}
\label{fig3}
\end{figure}
\begin{figure}
\caption{The overlap $M$ as a function 
of $\beta_{J}$ for several values of $\Gamma$. 
We set the system parameters $T_{m}=T_{s}=0.9, 
\tau=\tau_{0}=1.0,J=J_{0}=1.0$ and 
$h/\beta_m=0.9=\beta_\tau/\beta_s$.  
For the case of $\Gamma=0$, 
the optimal $\beta_{J}$ is naturally identical to 
$J_{0}/J^{2}=1.0$. 
As the quantum fluctuation $\Gamma$ increases, 
the overlap $M$ decreases because the quantum 
fluctuation destroys the recovered image.}
\label{fig4}
\end{figure}
\begin{figure}
\caption{The results by quantum Monte Carlo 
simulations for standard picture (
A Japanese {\it Kanji} stamp for the name of 
{\it suzuki} which is the most popular name in Japan. 
The size is $50\times50$.). 
From the upper left to the lower 
right, the original picture, the 
damaged picture, 
the results of $\Gamma=0.2, 
1.0, 1.7, 1.9, 2.1$ and $2.7$ are 
displayed. 
The noise rate is $10\%$ 
(The overlap between 
the original picture and 
the damaged one is $0.9$).
}
\label{fig5}
\end{figure}
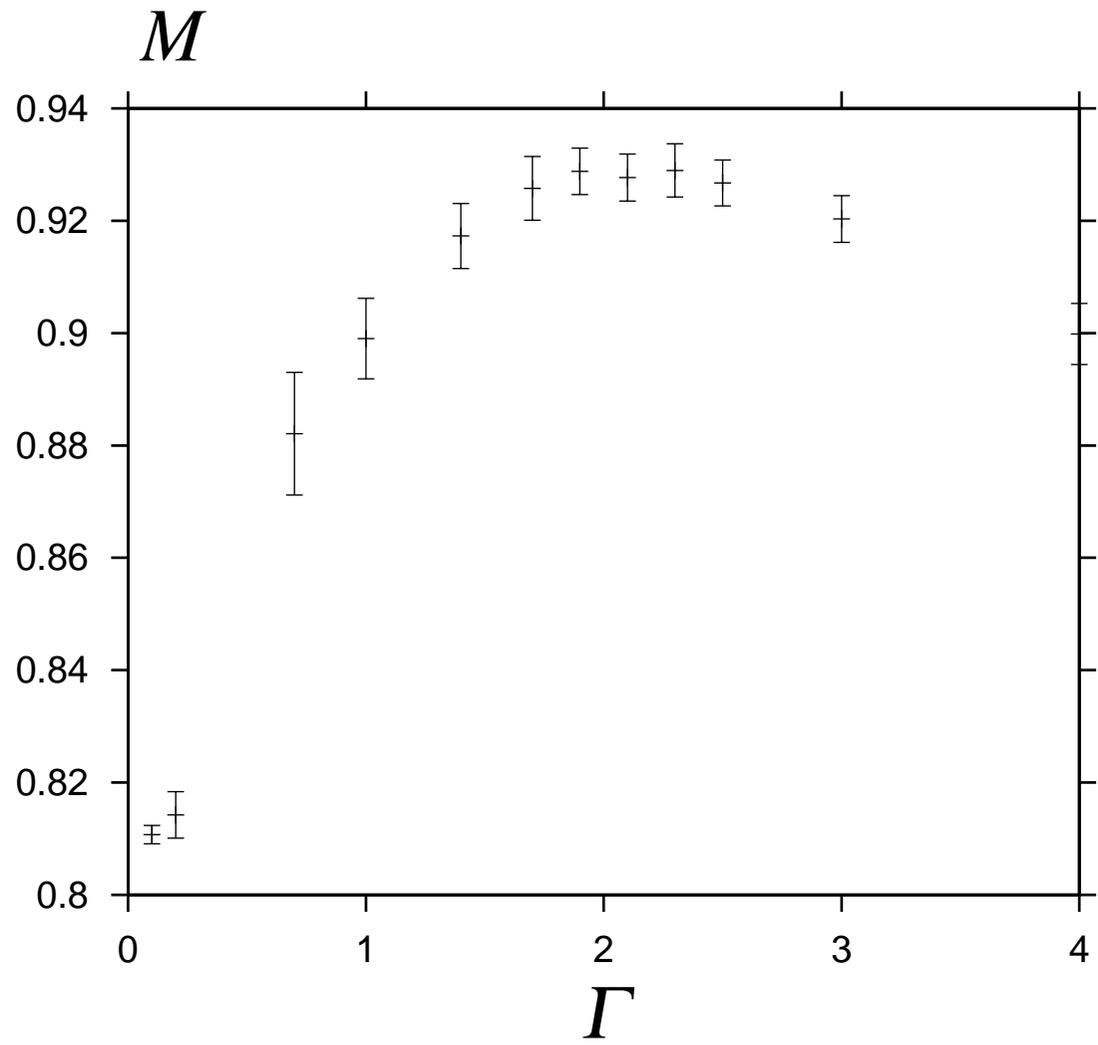
\begin{figure}
\caption{The overlap $M$ as a function of 
the quantum fluctuation $\Gamma$ for the standard 
picture in FIG. \ref{fig5}. 
We set $\beta_{m}=0.5, h=1.0$ and $P=50$. 
The error-bars are calculated by averaging over 
five independent runs. 
}
\label{fig6}
\end{figure}
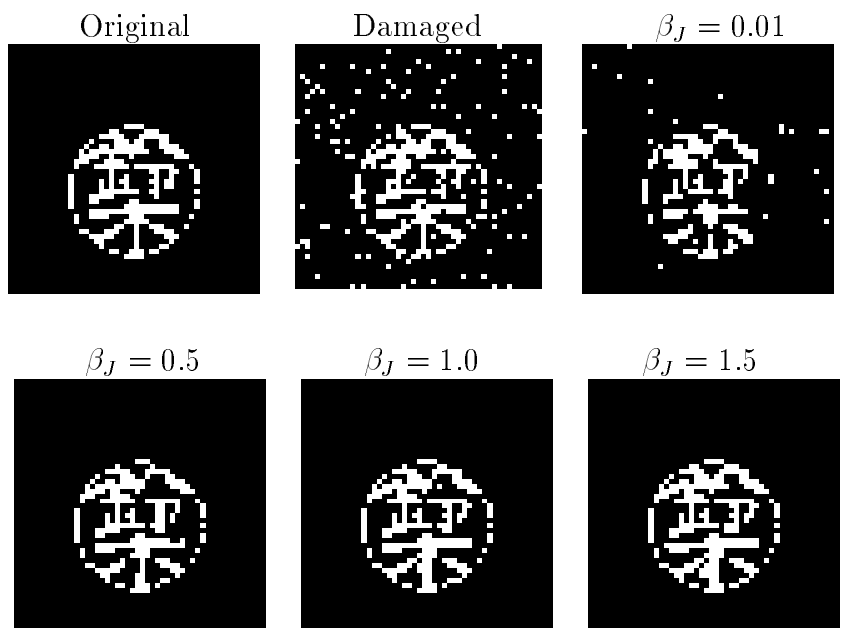
\begin{figure}
\caption{
The results by quantum Monte Carlo simulations 
including the parity check term. 
We fixed $\Gamma=2.0,h=1.0$ and 
$\beta_{m}=0.5$. 
From the upper left to the 
lower right, the original picture, 
the damaged picture, the results of 
$\beta_{J}=0.01,0.5,1.0$ and $1.5$ are 
shown. 
The noise rate is $10\%$ 
(The overlap between 
the original picture and 
the damaged one is $0.9$). 
}
\label{fig7}
\end{figure}
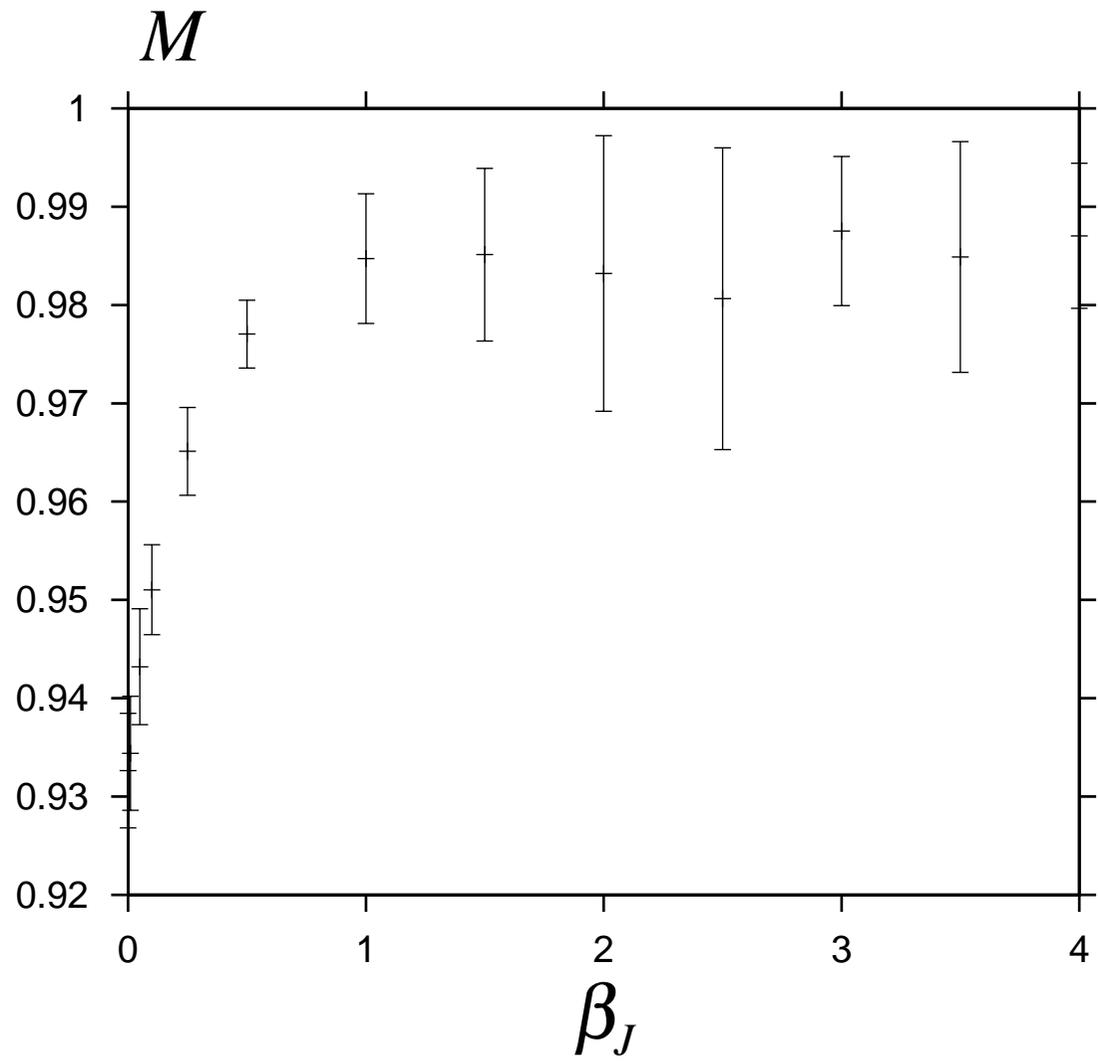
\begin{figure}
\caption{$\beta_{J}$-dependence 
of the overlap $M$ calculated by quantum Monte Carlo 
simulations for standard 
picture in FIG. \ref{fig6}. 
We set $\beta_{m}=0.5, h=1.0$ and 
$\Gamma=2.0$. 
The error-bars are calculated by 
five independent runs. 
}
\label{fig8}
\end{figure}
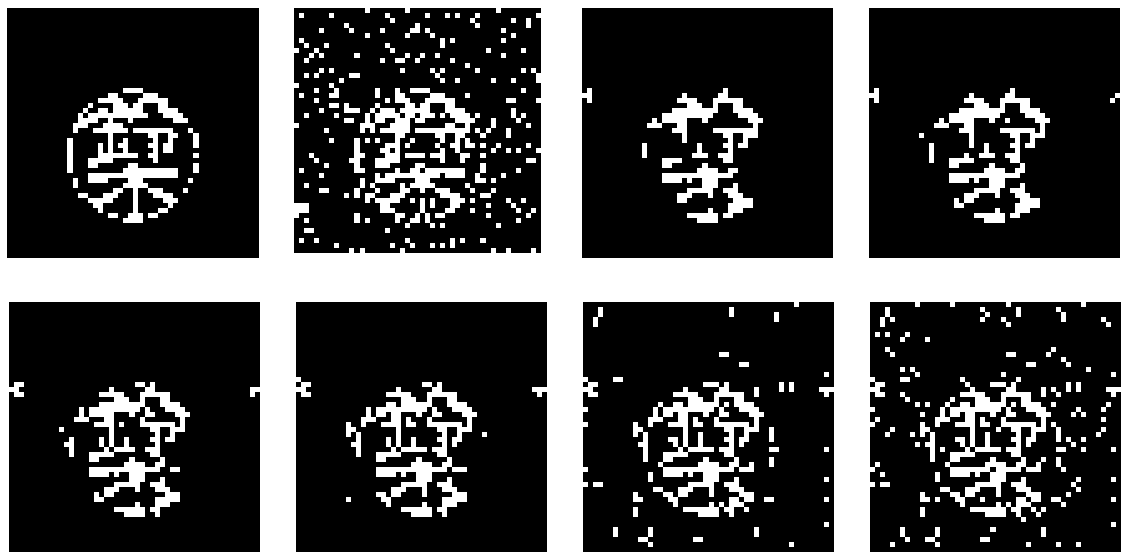
\begin{figure}
\caption{The restored pictures (
Their size are $50\times 50$) by 
quantum iterative algorithm for 
several values of $\Gamma$. 
From the upper left to 
the lower right, 
the original image,  
the corrupted image, 
the results of 
$\Gamma=0.001, 0.8, 1.2, 
1.6, 2.0$ and $\Gamma=3.0$. 
The noise rate is $20\%$ 
(The overlap between 
the original picture and 
the damaged one is $0.8$). 
}
\label{fig9}
\end{figure}
\begin{table}
\caption{The overlap $M$ calculated 
by the quantum iterative algorithm 
for several values of $\Gamma$. 
The restored pictures are 
shown in FIG. \ref{fig9}.   
The iteration times are also listed. 
We set $T=1, J=0.5$.}
\label{table1}
\end{table}

\begin{thebibliography}{9999}
\bibitem{Standard}
The standard pictures which are used in this paper 
are available at {\sf ftp://ftp.lab1.kuis.kyoto-u.ac.jp/pub/sidba/}.
\end{thebibliography}
\end{document}